\documentclass[12pt]{article}
\usepackage{amssymb}
\usepackage{amsmath}
\usepackage{graphicx}

\catcode`@=11
\renewcommand{\section}{\@startsection{section}{1}{0pt}{\medskipamount}
{\medskipamount}{\large\bf}}
\numberwithin{equation}{section}
\catcode`@=12

\def\beq{\begin{eqnarray}}    
\def\eeq{\end{eqnarray}}      

\def\ln{\,\mbox{ln}\,}                  

\def\im{\textrm{i}}

\def\sfrac#1#2{{\textstyle\frac{#1}{#2}}}
\def\={\ =\ }
\def\und{\qquad\textrm{and}\qquad}

\def\al{\alpha}

\def\de{\delta}

\def\vp{\varepsilon}

\begin{document}

\begin{titlepage}
\setcounter{page}{0}

\vskip 2.0cm

\begin{center}

{\LARGE\bf
Soft breaking of BRST symmetry and gauge dependence}

\vspace{18mm}

{\Large
P.M. Lavrov$\,{}^{\dagger}$, \
O.V. Radchenko$\,{}^{\dagger}$ \ and \
A.A. Reshetnyak$\,{}^\ast$
}

\vspace{8mm}

\noindent ${}^\dagger${\em
Tomsk State Pedagogical University,\\
Kievskaya St.\ 60, 634061 Tomsk, Russia}

\vspace{4mm}

$^\ast${\em
Institute of Strength Physics and Material Science, \\
Akademicheskii av.\ 2/4, 634021 Tomsk, Russia}

\vspace{18mm}

\begin{abstract}
\noindent We continue investigation of soft breaking of BRST
symmetry in the Batalin-Vilkovisky (BV) formalism beyond
regularizations like dimensional ones used in our previous paper
[JHEP 1110 (2011) 043, arXiv:1108.4820 [hep-th]]. We generalize a
definition of soft breaking of BRST symmetry valid for general
gauge theories and arbitrary gauge fixing. The gauge dependence of
generating functionals of Green's functions is investigated. It is
proved that such introduction of a soft breaking of BRST symmetry
into gauge theories leads to inconsistency of the conventional BV
formalism.
\end{abstract}

\end{center}

\vfill
\noindent{\sl Emails:} \ lavrov@tspu.edu.ru,
radchenko@tspu.edu.ru, reshet@ispms.tsc.ru


\end{titlepage}


\section{Introduction}\label{intro}

\noindent

A soft breaking of BRST symmetry in Yang-Mills theories has been
intensively studied in a series of recent papers \cite{Sorellas}.
This breakdown is connected with a restriction of the domain of
integration in the  functional integral due to the Gribov horizon
\cite{Gribov, Zwanziger1,Zwanziger2}. Note that all investigations
\cite{Sorellas,Zwanziger1,Zwanziger2} of the Gribov horizon in
Yang-Mills theories have been performed in the Landau gauge only.
Some months ago a generalization of notation of a soft breaking of
BRST symmetry for general gauge theories in arbitrary gauges in
the framework of BV formalism \cite{BV} has been proposed in
\cite{llr}. This approach to the problem of a soft breaking of
BRST symmetry includes Yang-Mills theories in the Landau gauge as
a very special case. In particular, it was shown that the way when
one takes into account the Gribov horizon in the form of an
additional term to the full action of given gauge system
coinciding with Gribov-Zwanziger action \cite{Gribov,
Zwanziger1,Zwanziger2} in case of Yang-Mills theories in the
Landau gauge leads to gauge dependent  S-matrix.

In the present paper we extend  the investigation performed in
\cite{llr} to case when one uses any regularization scheme
respecting gauge invariance.

The paper is organized as follows. In Section~\ref{sbBRST}, our
definition of the soft breaking of BRST symmetry is given in the
BV formalism. In Section~\ref{WI} we study different ways to
introduce BRST-like transformations  and  derive the Ward
identities for the generating functionals of Green's functions. In
Section~\ref{GD} we investigate the dependence of these
functionals on
 gauges. Concluding remarks are given in  Section~\ref{conclusion}.

We use the condensed notation of DeWitt \cite{DeWitt}. Derivatives
with respect to sources and antifields are taken from the left,
while those with respect to fields are taken from the right. Left
(right) derivatives with respect to fields (antifields) are
labelled by a subscript~$l$~($r$). The Grassmann parity of any
quantity $A$ in case of its homogeneity is denoted as $\varepsilon
(A)$.
\\

\section{Soft breaking of BRST symmetry in the BV
formalism}\label{sbBRST}

\noindent
Consider a theory of  gauge fields $A^i$, $i=1,2,\ldots,n$,
 ($\varepsilon(A^i)=\varepsilon_i$), with an initial
action $\mathcal{S}_0=\mathcal{S}_0(A)$. It is assumed the
invariance of this action under the gauge transformations $\delta
A^i= R^i_{\alpha}(A)\xi^{\alpha}$ with $\xi^{\alpha}$
($\varepsilon(\xi^{\alpha}) =\varepsilon_{\alpha}$) to be
arbitrary functions  of the space-time coordinates that means the
presence of the
 identities
\begin{eqnarray}
\label{GIClassA}   \mathcal{S}_{0,i}(A) R^i_{\alpha}(A)=0
\qquad\textrm{for}\quad \alpha=1,2,\ldots,m\ ,\quad 0<m<n\ .
\end{eqnarray}
among the classical extremals $\mathcal{S}_{0,i} = 0$. Here, the
functions $R^i_{\alpha}(A)$ ($\vp(R^i_\al)=\vp_i{+}\vp_\al$) are
the generators of the gauge transformations and we have used  the
 notation $\mathcal{S}_{0,i}\equiv\delta \mathcal{S}_0/\delta A^i$. The structure of configuration space $\{\Phi^A\}$ in
the  BV formalism depends on the type of given classical gauge
theory (for details, see \cite{BV}). For our purposes it is not
important to describe its explicit contents. We need in the fact
of existence of the total configuration space parameterized by the
fields $\Phi\ \equiv\ \{\Phi^A\}\=\{A^i,\ldots\}$ with
$\varepsilon(\Phi^A)=\varepsilon_A$ , where the dots indicate the
full set of additional to $A^i$ fields in the BV method. Then to
each field $\Phi^A$ of the configuration space, one should
introduce the corresponding antifield~$\Phi^*_A$ with  opposite
Grassmann parities to that of the corresponding field $\Phi^A$,
$\Phi^*\ \equiv\ \{\Phi^*_A\} \= \{A^*_i,\ldots\}$, with
$\vp(\Phi^*_A)=\vp_A{+}1$.

The main object of the BV quantization is a bosonic functional
 ${\bar S}={\bar S}(\Phi,\Phi^*)$
satisfying the master equation
\beq \label{MastEBV}
\sfrac {1}{2} ({\bar S},{\bar
S})\=\im\hbar\,{\Delta}{\bar S} \eeq
with the boundary condition
\beq \label{BoundCon} {\bar S}|_{\Phi^* = \hbar = 0}\=
\mathcal{S}_0(A)\ . \eeq
Here we used the notation of odd (i.e. $\vp
(\Delta)=1$) nilpotent  operator $\Delta$
\beq \label{DefAB}
\Delta\ \equiv\ (-1)^{\vp_A}
\frac{\delta_{\it l}} {\delta\Phi^A}\;\frac{\delta}
{\delta\Phi^*_A},\eeq
and the antibracket, $(F,G)$, which can be reproduced by  $\Delta$ acting
on the product of two functionals $F$ and $G$:
\beq
\Delta\;F \cdot G=(\Delta F)\cdot G+F\cdot (\Delta G)(-1)^{\varepsilon(F)}+
(F,G)^{\varepsilon(F)}.
\eeq
It is obvious that the initial action $\mathcal{S}_0$  satisfies
the classical master equation, $(\mathcal{S}_0,\mathcal{S}_0)=0$.

Having the action ${\bar S}$ and a corresponding fermionic gauge
fixing functional $\Psi = \Psi(\Phi)$, one can construct the
non-degenerate action $S_{ext}$ by the rule
\beq \label{ExtActBV}
S_{ext}(\Phi, \Phi^*) \= {\bar S}\big(
\Phi,\,\Phi^* + \sfrac{\de\Psi}{\de\Phi} \big)\ . \eeq
The action $S_{ext}$ satisfies the same master equation
(\ref{MastEBV}) as the functional ${\bar S}$, \beq
\label{ClMastEBVExt} \sfrac {1}{2} (S_{ext}, S_{ext})\=
\im\hbar\,{\Delta}{S_{ext}}\eeq and is used to construct the
generating functional of Green's functions in the BV
formalism~\cite{BV}.

Following to Refs. \cite{Zwanziger1, Zwanziger2, llr}, we deform
the action $S_{ext}$ by adding a functional $M=M(\Phi,\Phi^*)$,
thus determining  the full action $S$ as
\beq \label{Sfull} S\=S_{ext}+M\ .
\eeq
We shall speak of a \emph{soft breaking of BRST symmetry} in the
BV formalism if the condition %
\beq \label{SoftBrC}
\sfrac{1}{2}(M,M)\=-\im\hbar\,{\Delta}{M} \eeq
is fulfilled. Note,
that in classical limit,  $\hbar\rightarrow 0$, we assume that
$M=M_0+O(\hbar)$, Eq. (\ref{SoftBrC}) is reduced to \beq
\label{ClMasEqM} (M_0,M_0)=0. \eeq The reason to use the notation
of "a soft breaking of BRST symmetry`` may be explained as
follows. The master equation (\ref{ClMastEBVExt}) in the BV
formalism can be presented in the form
\beq
\Delta \exp \Big\{\frac{\im}{\hbar}S_{ext}\Big\}=0.
\label{MEexp}
\eeq
Using the action $S_{ext}$ as a solution to this equation in order
to construct Green's functions for general gauge theories one can
derive the BRST symmetry transformations \cite{BV}. Modifying the
action $S_{ext}$ by a {\it special} functional $M$ (it allows us
to speak of ''soft'') which satisfies the equation
\beq \label{SBRSTexp}\Delta \exp
\Big\{-\frac{\im}{\hbar}M\Big\}=0, \eeq
we obtain the action $S$ (\ref{Sfull})  not satisfying the equation
likes (\ref{MEexp})
\beq
\Delta \exp \Big\{\frac{\im}{\hbar}S\Big\}\neq 0.
\eeq
The BRST symmetry will be broken if we shall construct Green's
functions in the BV formalism using this action (see
Section~\ref{WI} for details). From (\ref{ClMastEBVExt}) and
(\ref{SoftBrC}) it follows that the basic equation of our approach
to the soft breaking of BRST symmetry reads %
\beq \label{CBasEq}
\sfrac{1}{2}(S,S)-\im\hbar\,{\Delta}{S}\=(S,M)\ .
\eeq
In classical limit, for $S=S_0+O(\hbar)$, it follows from
(\ref{CBasEq}) the equation,
\beq \label{CBasEq0}
\sfrac{1}{2}(S_0,S_0)\=(S_0,M_0),
\eeq
coinciding in classical limit,  $\hbar\rightarrow 0$, with the basic
equation to the soft breaking of BRST
symmetry considered in Ref.\cite{llr} when  a regularization likes
dimensional one  for the local functional $S$ is applied, $\Delta
S \sim \delta(0) = 0$.

It is important to  note that the condition (\ref{SoftBrC}) will be
automatically valid in case when  the soft breaking of BRST symmetry
originates from a modification of the integration measure  in the
path integral. In this case  $M$ will be a functional of the field
variables $\Phi^A$ only, i.e.~$M=M(\Phi)$. As we have already
mentioned in Ref. \cite{llr}, this is exactly the situation for
Yang-Mills theory in the Landau gauge, when one takes into account the
Gribov horizon~\cite{Zwanziger1,Zwanziger2}. We consider the more
general situation of $M=M(\Phi,\Phi^*)$  not restricting ourselves
to this special case.

Note that the right-hand side of the basic
 equation~(\ref{CBasEq}) can be presented in the form%
 \beq\label{brstop}
 (S,M)\={\hat s}M -\im\hbar\,{\Delta}M\ ,\eeq%
  where ${\hat s}$
denotes the quantum Slavnov-Taylor operator defined as
$\hbar$-deformation of its classical analog in the form
 \beq {\hat
s}\=(S_{ext},\bullet)-\im\hbar\,{\Delta}\ .
\eeq
Due to master equation for $S_{ext}$ (\ref{ClMastEBVExt}) this operator is
nilpotent,
\beq
{\hat s}^2\=0\ .
\eeq
However, as compared to the
consideration in \cite{llr}, the presence of the additional term to
${\hat s}$ in the right-hand side of the relation
(\ref{brstop}) leads to the inequality \beq {\hat
s}\,\Big\{\sfrac{1}{2}(S,S)-\im\hbar \Delta S\Big\}\ \ne 0\ . \eeq
which being written for the action $S_{ext}$ should be the identity
for general gauge theories {\it without\/} a (soft) breaking of BRST symmetry.

\section{BRST transformations and Ward identities}\label{WI}

\noindent Let us consider some quantum consequences of the
classical equations (\ref{ClMastEBVExt}), (\ref{SoftBrC})
and~(\ref{CBasEq}). To this end we introduce the generating
functional of Green's functions,
\beq \label{ZBV} Z(J,\Phi^*)\=\int\!D\Phi\ \exp
\Big\{\frac{\im}{\hbar} \big(S(\Phi,\Phi^*)+ J_A\Phi^A\big)\Big\}\ ,
\eeq
where $S(\Phi,\Phi^*)$ satisfies the basic classical equation
(\ref{CBasEq}) and has the form~(\ref{Sfull}). Furthermore, $J_A$
are the usual external sources for the fields~$\Phi^A$. The
Grassmann parities of these sources are defined in a natural way, \
$\vp(J_A) = \vp_A$.

Let us consider  the vacuum functional $Z(0,\Phi^*) = Z(\Phi^*)$
and its integrand
\beq \label{intgM}
{\cal M}={\cal M}(\Phi,\Phi^*)=D\Phi\ \exp
\Big\{\frac{\im}{\hbar} S(\Phi,\Phi^*)\Big\}. \eeq
In the BV formalism the BRST symmetry appears as invariance of the
integrand of vacuum functional under the change of variables,
$\Phi^A \to \Phi^{\prime A} = \Phi^A +\delta_B \Phi^A  $
determined with the help of $S_{ext}(\Phi,\Phi^*)$,
\beq \label{BRST}
\delta_B \Phi^A = \frac{\delta  S_{ext}}{\delta
\Phi^*_A}\theta, \qquad  \delta_B \Phi^*_A = 0,
\eeq
where $\theta$ is a nilpotent constant odd parameter. Carrying out
the change of variables (\ref{BRST}) in (\ref{intgM}) we obtain
\beq \label{Zvacnbrst} {\cal M}^{'}= {\cal M}\Big(1 -
\frac{\im}{\hbar}\theta \;\frac{\delta M}{\delta \Phi^A}\frac{\delta
S_{ext}}{\delta \Phi^*_A}  \Big). \eeq
Non-invariance of the integrand means violation of the standard
BRST symmetry. Of course, one may think that it is necessary to
modify the definition of BRST transformations in case of the
theory under consideration. A possible way for its modification
looks as
\beq
\label{modBRST}
\delta_B \Phi^A = \frac{\delta  S}{\delta
\Phi^*_A}\theta, \qquad  \delta_B \Phi^*_A = 0,
\eeq
Performing the change of variables (\ref{modBRST})  in the
integrand functional ${\cal M}$ (\ref{intgM}) one has
\beq \label{Zvacnmodbrst} {\cal M}^{'}= {\cal M}\Big(1 -
\frac{\im}{\hbar}\theta \;(S,M)  \Big). \eeq
We meet again the non-invariance under modified BRST
transformations (\ref{modBRST}). Moreover one can consider the
following one-parameter functional $S_{\kappa}$
\beq
S_{\kappa}=S_{ext}+\kappa M
\eeq
to define the BRST trasformations
\beq
\label{nmodBRST}
\delta_{B_{\kappa}} \Phi^A = \frac{\delta  S_{\kappa}}{\delta
\Phi^*_A}\theta, \qquad  \delta_{B_{\kappa}} \Phi^*_A = 0.
\eeq
Here $\kappa$ is a real number. We have  the standard BRST
transformations (\ref{BRST}) for $\kappa=0$ and the modified ones
(\ref{modBRST}) for $\kappa=1$. Carrying out the change of
variables (\ref{nmodBRST}) in the integrand (\ref{intgM})
 we arrive at the result
\beq \label{ZBVnmodbrst} {\cal M}^{'}= {\cal M}\Big(1 -
\frac{\im}{\hbar}\theta \; \Big[\kappa (S_{ext}, M)\ -2
\im\hbar\,\kappa {\Delta}M + (1-\kappa)\frac{\delta M}{\delta
\Phi^A}\frac{\delta S_{ext}}{\delta \Phi^*_A} \Big]\Big ) . \eeq
We may conclude that the non-invariance of the integrand exists
for any choice of the para\-meter~$\kappa$.

Turning to the properties of the generating functional
$Z(J,\Phi^*)$ note, that from the Eq. (\ref{MEexp}) it follows
that its averaging over the total configuration space of the
fields
 $\Phi^A$ with measure $\exp
\big\{\frac{\im}{\hbar} \big(S+J_A\Phi^A\big)\big\}$
\beq \nonumber 0 \= \int\!D\Phi\ \Delta \exp
\Big\{\frac{\im}{\hbar}S_{ext}\Big\}\; \exp \Big\{\frac{\im}{\hbar}
\big(S+J_A\Phi^A\big)\Big\}\ \eeq
leads after integrating by parts in the functional integral to the
following identity for the generating functional $Z$,
\beq
\label{WIZBV}
&&\Big(J_A+M_{A}\big(\sfrac{\hbar}{\im}\sfrac{\delta}{\delta
J},\Phi^*\big)\Big)\left(\frac{\hbar}{\im}\frac{\delta
}{\delta\Phi^*_A}\ -\
M^{A*}\big(\sfrac{\hbar}{\im}\sfrac{\delta}{\delta
J},\Phi^*\big)\right)Z(J,\Phi^*)=0.
\eeq
Here the notations
\beq \nonumber
M_{A}\big(\sfrac{\hbar}{\im}\sfrac{\delta}{\delta
J},\Phi^*\big)\equiv \frac{\delta M(\Phi,\Phi^*)}{\delta
\Phi^A}\Big|_{\Phi\rightarrow
\frac{\hbar}{\im}\frac{\delta}{\delta J}} \und
M^{A*}\big(\sfrac{\hbar}{\im}\sfrac{\delta}{\delta
J},\Phi^*\big)\equiv \frac{\delta
M(\Phi,\Phi^*)}{\delta\Phi^*_A}\Big|_{\Phi\rightarrow
\frac{\hbar}{\im}\frac{\delta}{\delta J}}
\eeq
have been used. In case of $M=0$, the identity  (\ref{WIZBV}) is
reduced to the usual Ward identity for the generating functional of
Green's functions in the BV formalism. Hence, we refer
to~(\ref{WIZBV}) as the Ward identity for~$Z$ in a gauge theory with
softly broken BRST symmetry. Note, that for the regularization
scheme likes dimensional one, we will have $\Delta M = 0$ and,
therefore the equation (\ref{SoftBrC}) is reduced to
$\big(M,\,M\big)=0$, which leads to the vanishing of the
combination, $M_{A}\big(\sfrac{\hbar}{\im}\sfrac{\delta}{\delta
J},\Phi^*\big) M^{A*}\big(\sfrac{\hbar}{\im}\sfrac{\delta}{\delta
J},\Phi^*\big)$ in the Ward identity (\ref{WIZBV}) as it was derived
in \cite{llr}.

Then, introducing the generating functional of connected Green's
functions, %
\beq W(J,\Phi^*) \= -\im\hbar \ln Z(J,\Phi^*)\ ,%
\eeq
the identity (\ref{WIZBV}) can be rewritten for $W$ as %
\beq \label{WIWBV}
 &&\Big(J_A+M_{A}\big(\sfrac{\delta W}{\delta J}+
\sfrac{\hbar}{\im}\sfrac{\delta}{\delta
J},\Phi^*\big)\Big)\left(\frac{\delta
W(J,\Phi^*)}{\delta\Phi^*_A}\ -\ M^{A*}\big(\sfrac{\delta
W}{\delta J}+\sfrac{\hbar}{\im}\sfrac{\delta}{\delta
J},\Phi^*\big)\right)=0 . \eeq

The generating functional of the vertex functions (or effective
action) is obtained by Legendre transforming of $W$, \beq
\label{EA} \Gamma (\Phi,\,\Phi^*) \= W(J,\Phi^*) - J_A\Phi^A,
\qquad\textrm{where}\quad \Phi^A = \frac{\delta W}{\delta J_A},
\qquad \frac{\delta\Gamma}{\delta\Phi^A}=-J_A\ . \eeq Taking into
account the equality, $\frac{\delta \Gamma}{\delta
\Phi^*_A}\=\frac{\delta W}{\delta\Phi^*_A}$ ,
 we can rewrite the identity
(\ref{WIWBV}) in terms of~$\Gamma$ as
\beq \label{WIGammaBV}
\sfrac{1}{2}(\Gamma,\Gamma) \=
\frac{\delta\Gamma}{\delta\Phi^A}{\widehat M}^{A*}
+{\widehat M}_{A}\frac{\delta \Gamma}{\delta
\Phi^*_A}-{\widehat M}_{A}{\widehat M}^{A*}\ .
\eeq
Here, we have used the notations \beq {\widehat M}_{A}\ \equiv\
\frac{\delta
M(\Phi,\Phi^*)}{\delta\Phi^A}\Big|_{\Phi\to\widehat\Phi} \und
{\widehat M}^{A*}\ \equiv\ \frac{\delta
M(\Phi,\Phi^*)}{\delta\Phi^*_A}\Big|_{\Phi\to\widehat\Phi}\,, \eeq
where the sign ${\widehat\Phi}^A$ means the field $\Phi^A$
enlarged  by the derivatives $\frac{\delta_l}{\delta\Phi^B}$, %
 \beq
{\widehat\Phi}^A\=\Phi^A+\im\hbar\,(\Gamma^{''-1})^{AB}
\frac{\delta_l}{\delta\Phi^B} \eeq and the matrix
$(\Gamma^{''-1})$ is inverse to the matrix $\Gamma^{''}$ with
elements \beq (\Gamma^{''})_{AB}\=\frac{\delta_l}{\delta\Phi^A}
\Big(\frac{\delta\Gamma}{\delta\Phi^B}\Big)\
,\qquad\textrm{i.e.}\quad
(\Gamma^{''-1})^{AC}(\Gamma^{''})_{CB}=\delta^A_{\ B}\ . \eeq We
see again, in the case $M=0$ the identity (\ref{WIGammaBV})
coincides with the Ward identity for the effective action in the
BV formalism. Emphasize that the identity (\ref{WIGammaBV}) is
compatible with the classical equation (\ref{CBasEq}), since
$\hbar\to0$ yields $\Gamma=S_0$, ${\widehat M}=M_0$, and
(\ref{WIGammaBV})  is reduced to~(\ref{CBasEq0}).

In similar manner  we can derive the Ward identity
which follows from (\ref{SoftBrC}). To this end, we average the
equation (\ref{SoftBrC})  over the  configuration space of the
fields
 $\Phi^A$ with measure $\exp
\big\{\frac{\im}{\hbar} \big(S+J_A\Phi^A\big)\big\}$,
 \beq
\nonumber 0 \= \int\!D\Phi\ \Big\{\frac{\delta
M}{\delta\Phi^A}\frac{\delta M} {\delta\Phi^*_A}+
\im\hbar\,{\Delta}{M}\Big\}\; \exp \Big\{\frac{\im}{\hbar}
\big(S+J_A\Phi^A\big)\Big\}\ . \eeq
and derive after usual manipulations with functional integral the
identity in terms of mean fields $\Phi^A$ (\ref{EA}) %
 \beq \label{WIM}{\widehat M}_{A}{\widehat M}^{A*}=-i\hbar {\widehat
M}_A^{\;\;A^*} \eeq
where the notation
\beq
{\widehat M}_A^{\;\;A^*}=
\frac{\delta^2 M}{\delta\Phi^*_A\delta\Phi^A}\Big|_{\Phi\to\widehat\Phi}
\eeq
was used. The identity (\ref{WIM})  is reduced to the known
identity, ${\widehat M}_{A}{\widehat M}^{A*}=0$, derived in
\cite{llr}, when the regularization scheme likes dimensional one
is applied.
\\

\section{Gauge dependence}\label{GD}

\noindent Here we  study the gauge dependence of the generating
functionals $Z$, $W$ and $\Gamma$ for general gauge theories with
a soft breaking of BRST symmetry as it was defined in the section
above. The derivation of this dependence is based on the fact that
any variation of the gauge-fixing functional,
$\Psi(\Phi)\rightarrow\Psi(\Phi)+\delta\Psi(\Phi)$, leads to a
variation both the action $S_{ext}$ (\ref{ExtActBV}), the
functional $Z$~\cite{VLT} and the functional $M$. The variation of
$S_{ext}$ can be presented in the form \beq \label{varSext} \delta
S_{ext}\=\frac{\delta \delta\Psi}{\delta\Phi^A}\,\frac{\delta
S_{ext}}{\delta\Phi^*_A}\qquad \texttt{ or as } \qquad \delta
S_{ext}\=-(S_{ext},\delta\Psi)\= -{\hat s}\,\delta\Psi\ , \eeq
 whereas the variation of $M$ for  the variation~$\delta\Psi$
 we denote as $\delta M(\Phi,\Phi^*)$.
 From (\ref{ZBV}), (\ref{varSext}) and the variation of $M$
we obtain the gauge variation of~$Z$,
\beq \label{varZ}
\delta Z(J,\Phi^*)\=\frac{\im}{\hbar}\int\!D\Phi\ \Big( \frac{\delta
\delta\Psi}{\delta\Phi^A}\frac{\delta S_{ext}}{\delta\Phi^*_A}+\delta
M\Big)\; \exp \Big\{\frac{\im}{\hbar}
\big(S(\Phi,\Phi^*)+ J_A\Phi^A\big)\Big\}\ .
\eeq
Then applying the equality
\beq \label{AuxId} \nonumber
0&=&\int\!D\Phi\ \frac{\delta_l}{\delta
\Phi^A}\Big[\delta\Psi\;\frac{\delta S_{ext}}{\delta\Phi^*_A}\;\exp
\Big\{\frac{\im}{\hbar} \big(S(\Phi,\Phi^*)+
J_A\Phi^A\big)\Big\}\Big] \\ \nonumber
&=&\int\!D\Phi\ \Big[\frac{\delta\delta\Psi
}{\delta\Phi^A}\,\frac{\delta
S_{ext}}{\delta\Phi^*_A}-\frac{\im}{\hbar}\Big(J_A+\frac{\delta
S}{\delta \Phi^A}\Big)\frac{\delta
S_{ext}}{\delta\Phi^*_A}\,\delta\Psi\Big]\exp \Big\{\frac{\im}{\hbar}
\big(S(\Phi,\Phi^*)+ J_A\Phi^A\big)\Big\}\ ,
\eeq
for which validity  the equation (\ref{ClMastEBVExt}) was used,
we can rewrite (\ref{varZ}) as
\beq \nonumber \delta Z(J,\Phi^*) &=&
\frac{\im}{\hbar}\Big[\Big(J_A+M_{A}
\big(\sfrac{\hbar}{\im}\sfrac{\delta}{\delta
J},\Phi^*\big)\Big)\left(\frac{\delta}{\delta\Phi^*_A}\,
 -\frac{\im}{\hbar}
M^{A*} \big(\sfrac{\hbar}{\im}\sfrac{\delta}{\delta
J},\Phi^*\big)\right)\,\delta\Psi
\big(\sfrac{\hbar}{\im}\sfrac{\delta}{\delta J}\big) \\
&& + \delta M\big(\sfrac{\hbar}{\im}\sfrac{\delta}{\delta
J},\Phi^*\big)\Big]Z(J,\Phi^*)\label{varZ1}.
\eeq

In turn, the corresponding variation of the generating
functional of connected Green's functions, $\delta W(J,\Phi^*)\=
\sfrac{\hbar}{\im}Z^{-1}\delta Z$, takes the form
\beq
\label{varW} \delta W(J,\Phi^*)=
\Big(J_A+M_{A}\big(\sfrac{\delta W}{\delta
J}+\sfrac{\hbar}{\im}\sfrac{\delta}{\delta
J},\Phi^*\big)\Big)\frac{\delta }{\delta\Phi^*_A} \delta\Psi
\big(\sfrac{\delta W}{\delta
J}+\sfrac{\hbar}{\im}\sfrac{\delta}{\delta J}\big) \ +\ \delta
M\big(\sfrac{\delta W}{\delta
J}+\sfrac{\hbar}{\im}\sfrac{\delta}{\delta J},\Phi^*\big) \,,
\eeq
with using the Ward identity (\ref{WIWBV}).

Now, we are able to arrive at  our final purpose concerning  the
derivation of the gauge variation of the effective action.  First,
we note that $\delta \Gamma=\delta W$. Second, we observe that the
change of the variables $(J_A, \Phi^*_A) \to (\Phi^A, \Phi^*_A)$
from the Legendre transformation  (\ref{EA}) implies that
\beq
\label{dphistar}
{\frac{\delta}{\delta\Phi^*}}\Big|_{J}=
\frac{\delta}{\delta\Phi^*}\Big|_{\Phi} + \frac{\delta
\Phi}{\delta\Phi^*}{\frac{\delta_{\it
l}}{\delta\Phi}}\Big|_{\Phi^*}.
\eeq

Next, differentiating  the Ward identities for
$Z$~(\ref{WIZBV}) with respect to the sources $J_B$,  we obtain
\beq \nonumber
 &&\frac{\hbar}{i}\frac{\delta Z}{\delta
\Phi^*_B}+\frac{\hbar}{i}
\Big(J_A+M_{A}\big(\sfrac{\hbar}{\im}\sfrac{\delta}{\delta
J},\Phi^*\big) \frac{\delta^2 Z}{\delta
J_B\delta\Phi^*_A}(-1)^{\varepsilon_A\varepsilon_B}
 -M^{B*}\big(\sfrac{\hbar}{\im}\sfrac{\delta}{\delta
 J},\Phi^*\big)Z-\\
&&-(-1)^{\varepsilon_B}J_AM^{A*}
\big(\sfrac{\hbar}{\im}\sfrac{\delta}{\delta
J},\Phi^*\big)\frac{\delta Z}{\delta J_B}-
(-1)^{\varepsilon_B}M_{A}\big(\sfrac{\hbar}{\im}\sfrac{\delta}{\delta
J},\Phi^*\big)M^{A*} \big(\sfrac{\hbar}{\im}\sfrac{\delta}{\delta
J},\Phi^*\big)\frac{\delta Z}{\delta J_B}=0. \label{dWIZ}
\eeq
Using the interrelation of the  derivatives for $Z$ and $W$,
\beq && \left(\frac{\delta Z}{\delta J_A}, \frac{\delta Z}{\delta
\Phi^*_A}\right) = \frac{\im}{\hbar}\exp\{\sfrac{\im}{\hbar} W\}
\left(\frac{\delta W}{\delta J_A},  \frac{\delta W}{\delta
\Phi^*_A}\right),\\
&&   \frac{\delta^2 Z}{\delta \Phi^*_B\delta J_A}
=\exp\{\sfrac{\im}{\hbar} W\}\left[
\left(\frac{\im}{\hbar}\right)^2 \frac{\delta W}{\delta
\Phi^*_B}\frac{\delta W}{\delta
J_A}+\frac{\im}{\hbar}\frac{\delta^2 W}{\delta \Phi^*_B\delta
J_A}\right],\eeq
the Eqs. (\ref{dWIZ}) may be written in terms of functional $W$
as,
\beq \label{dW} &&  \frac{\delta
W(J,\Phi^*)}{\delta\Phi^*_B}+(-1)^{\vp_B}
\Big(J_A+M_{A}\big(\sfrac{\hbar}{\im}\sfrac{\delta}{\delta
J}+\sfrac{\delta W}{\delta J},\Phi^*\big)\Big)\frac{\delta^2
W(J,\Phi^*)}{\delta\Phi^*_A \delta J_B}\ -\
M^{B*}\big(\sfrac{\hbar}{\im}\sfrac{\delta}{\delta
J}+\sfrac{\delta W}{\delta
J},\Phi^*\big)\nonumber\\
&&  \= -\frac{\im}{\hbar}(-1)^{\vp_B}
\Big(J_A+M_{A}\big(\sfrac{\hbar}{\im}\sfrac{\delta}{\delta
J}+\sfrac{\delta W}{\delta J},\Phi^*\big)\Big) \left(\frac{\delta
W}{\delta
\Phi^*_A}-M^{A*}\big(\sfrac{\hbar}{\im}\sfrac{\delta}{\delta
J}+\sfrac{\delta W}{\delta J},\Phi^*\big)\right)\frac{\delta
W}{\delta J_B}\ . \eeq
From (\ref{dW}) it follows
\beq  \frac{\delta \Gamma}{\delta\Phi^*_B}- {\widehat M}^{B^*}
-\Big(\frac{\delta\Gamma}{\delta\Phi^A}-{\widehat
M}_{A}\Big)\frac{\delta\Phi^B}{\delta\Phi^*_A}
(-1)^{\varepsilon_B}  = \frac{i}{\hbar}(-1)^{\varepsilon_B}
\Big(\frac{\delta\Gamma}{\delta\Phi^A}-{\widehat
M}_{A}\Big)\Big(\frac{\delta \Gamma}{\delta\Phi^*_A}- {\widehat
M}^{A^*}\Big)\Phi^B. \label{dG} \eeq
To simplify the above expression one should commute the fields
$\Phi^B$ to the left in the last summand in order  to use Ward
identity for effective action $\Gamma$
(\ref{WIGammaBV})
%
As a result, we rewrite the relation (\ref{dG}) in the form
\beq \nonumber -\Big(\frac{\delta\Gamma}{\delta\Phi^A}-{\widehat
M}_{A}\Big)\frac{\delta\Phi^B}{\delta\Phi^*_A}&=&
-\Big(\frac{\delta\Gamma}{\delta\Phi^*_B}-{\widehat
M}^{B^*}\Big)(-1)^{\varepsilon_B}\\\label{dGamma}
&&+\frac{i}{\hbar}\Big[-{\widehat
M}_{A}\frac{\delta\Gamma}{\delta\Phi^*_A}-
\frac{\delta\Gamma}{\delta\Phi^A}{\widehat M}^{A^*}+{\widehat
M}_{A}{\widehat M}^{A^*},\Phi^B\Big], \eeq
where the square brackets $\big[\ ,\ \big]$ denote  the
supercommutator.

From (\ref{varW}), (\ref{dphistar}) and (\ref{dGamma})
the variation of the effective action can be presented in the form,
\beq \delta\Gamma &=& -(\Gamma,\langle\delta\Psi\rangle)\ +\
\left({\widehat M}_{A} \frac{\delta}{\delta \Phi^{*}_{A}} +\
(-1)^{\vp_A} {\widehat
M}^{A*} \frac{\delta_l}{\delta \Phi^A}\right)\langle\delta\Psi\rangle  \nonumber \\
&& -\ \frac{\im}{\hbar}\Big[{\widehat M}_A \frac
{\delta\Gamma}{\delta\Phi^*_A}+\frac{\delta\Gamma}{\delta
\Phi^A}{\widehat M}^{A*} -{\widehat M}_A{\widehat M}^{A*}  ,\
\Phi^B\Big] \frac{\delta_{\it
l}}{\delta\Phi^B}\,\langle\delta\Psi\rangle\ +\ \langle\delta
M\rangle \label{varGamma}\ , \eeq
with local (for $M=0$) operator acting on the functional $
\langle\delta\Psi\rangle$. Here  we imply the notations
\beq \langle\delta\Psi\rangle\=\delta\Psi({\widehat\Phi})\cdot 1
\und \langle\delta M\rangle\=\delta M({\widehat \Phi},\Phi^*)\cdot
1 \ .\eeq
Then, using  the  identities,
\beq \frac{\delta \Phi^B}{\delta\Phi^*_A} =
(-1)^{\vp_B(\vp_A+1)}\frac{\delta }{\delta J_B} \frac{\delta W}{
\delta \Phi^*_A} =- (-1)^{\vp_B(\vp_A+1)}(\Gamma^{''-1})^{BC}
\frac{\delta_{\it l} }{\delta \Phi^C}\frac{\delta \Gamma}{ \delta
\Phi^*_A}, \eeq
following from the Legendre transformation (\ref{EA}) we can
present the variation of the effective action in the equivalent,
so-called non-local (due to explicit presence of the quantities
$(\Gamma^{''-1})^{BC}$) form,
 \beq \delta\Gamma \= \frac{\delta\Gamma}{\delta\Phi^A}
{\widehat F}^A\,\langle\delta\Psi\rangle\ -\ {\widehat
M}_A{\widehat F}^A \langle\delta\Psi\rangle\ +\ \langle\delta
M\rangle\ , \label{varGammaF} \eeq
where the operator  ${\widehat F}^A$ is derived from the Eqs.
(\ref{dphistar}), (\ref{dGamma}), (\ref{varGamma}) as follows
\beq {\widehat F}^A &=&-\frac{\delta}{\delta\Phi^*_A}\ +\
(-1)^{\vp_B(\vp_A+1)} (\Gamma^{''-1})^{BC}\Big(\frac{\delta_{\it
l}}{\delta\Phi^C}\frac {\delta
\Gamma}{\delta\Phi^{*}_{A}}\Big)\frac{\delta_{\it l}
}{\delta\Phi^B}\ . \label{FAdef} \eeq

From the variation (\ref{varGammaF}) it follows that on shell the
effective action is generally gauge dependent because of
\beq \frac{\delta\Gamma}{\delta\Phi^A}=0 \qquad
\longrightarrow\qquad \delta\Gamma\neq 0\ .
\eeq
This fact does not permit to formulate  consistently   a soft
breaking of BRST symmetry within the field-antifield formalism, if
only  two last terms in~(\ref{varGammaF}) cancel each other,
\beq \label{BasRest}
\langle\delta M\rangle\={\widehat M}_A{\widehat F}^A \langle\delta\Psi\rangle\ .
\eeq
However, this is rather a strong restriction on the BRST-breaking
functional~$M$ for the effective action to be gauge independent
on-shell. The same statement is valid for the physical S-matrix.
Really, Eq. (\ref{BasRest})~fixes the gauge variation of
$M=M(\Phi,\Phi^*)$ under a change of the gauge-fixing
functional~$\Psi$ to be
\beq \label{BEqvM} \delta M\=\frac{\delta
M}{\delta\Phi^A}\,{\widehat F}_0^A\,\delta\Psi\quad\texttt{ where
}\quad {\widehat F}_0^A\=(-1)^{\vp_B(\vp_A+1) } (S^{''-1})^{BC}
\Big(\frac{\delta_{\it l}}{\delta\Phi^C}\frac {\delta
S}{\delta\Phi^{*}_{A}}\Big) \frac{\delta_{\it l} }{\delta\Phi^B}\
. \eeq
It was shown in \cite{llr} that already in the case of Yang-Mills
theories in linear $R_{\xi}$ gauge which includes the Landau gauge
the relation (\ref{BEqvM}) does not satisfy. We are forced to claim
that a consistent quantization of general gauge theories when
restriction on the domain of integration in functional integral  is
taken as an addition to the full action of a given gauge system
violating the BRST symmetry does not exist.

\section{Conclusions}\label{conclusion}

\noindent In the given paper we have suggested a definition of
soft breaking of BRST symmetry in the BV formalism using any
regularization scheme respecting gauge invariance. To this
purpose, we added a `breaking functional'~$M$  to the gauge-fixed
action $S_{ext}$ which, in turn,  is constructed from an arbitrary
classical gauge-invariant action $\mathcal{S}_0$ according to the
rules of the BV method.  The soft breaking of BRST symmetry was
determined by the analog of the quantum master equation,
$(M,M)=-2i\hbar \Delta M$. It was proved the non-invariance of the
integrand of
 vacuum functional under the  BRST transformations determined
by means of the functional, $(S_{ext}+\kappa M)$, for any value of
the real parameter $\kappa$.
 We have obtained all Ward identities for the generating
functional of Green's functions $Z$, of connected Green's
functions $W$ and of vertex functions $\Gamma$ being different
from ones in Ref.\cite{llr} due to non-vanishing, in general, of
the operator $\Delta$ action on the functionals $S_{ext}$ and $M$.
The Ward  identities were used to investigate the gauge dependence
of those functionals. It was argued that $\Gamma$ as well as the
S-matrix are on-shell gauge dependent. We were forced to claim
that a consistent quantization of gauge systems in the BV
formalism with the soft breaking of BRST symmetry does not exist.

\section*{Acknowledgments}
\noindent The work is supported by  the LRSS grant 224.2012.2  as
well as by the RFBR grant 12-02-00121 and the RFBR-Ukraine grant
11-02-90445.

\end{document}